# Ionic Disorder-Mediated Exfoliation and Optical Birefringence in a Non-van der Waals Oxide

*Nikolay V. Pak[1,2], Mikhail K. Tatmyshevskiy[2], Ivan A. Kruglov[1,2], Konstantin V. Kravtsov[1,2], Anton A. Minnekhanov[1], Dmitriy V. Grudinin[1], Adilet N. Toksumakov[1], Aleksandr S. Slavich[1], Dmitry I. Yakubovsky[2], Andrey A. Vyshnevyy[1,2], Marwa A. El-Sayed[2,3], Georgy A. Ermolaev[1], Aleksey V. Arsenin[1,2] and Valentyn S. Volkov[1,a]*

[1] *Emerging Technologies Research Center, XPANCEO, Internet City, Emmay Tower, Dubai, United Arab Emirates*
[2] *Moscow Center for Advanced Studies, Kulakova str. 20, Moscow, 123592, Russia*
[3] *Department of Physics, Faculty of Science, Menoufia University, Shebin El-Koom, 32511, Egypt*

[a] *Author to whom correspondence should be addressed: vsv@xpanceo.com*

## Abstract

The landscape of two-dimensional photonics has been dominated by van der Waals (vdW) materials. Expanding this library to include non-vdW layered systems promises enhanced environmental robustness and access to novel functionalities, such as strong ionic conductivity, yet their exfoliation remains challenging. Here, we establish $Na_2Zn_2TeO_6$ (NZTO), a P2-type superionic conductor, as an exfoliable non-vdW optical material. We demonstrate that the highly disordered, mobile $Na^+$ interlayer inherently facilitates mechanical cleavage down to few-nanometer thicknesses (≈ 4 nm). Optical interrogation via spectroscopic ellipsometry reveals NZTO as a wide-bandgap dielectric with pronounced optical birefringence ($\Delta n$ ≈ 0.25) across the visible and near-infrared spectrum. The lattice dynamics, probed by temperature-resolved Raman spectroscopy, underscore the rigidity of the $[Zn_2TeO_6]^{2-}$ framework, which remains largely decoupled from the high ionic mobility. These results identify NZTO as a compelling platform for robust, anisotropic dielectric photonics, simultaneously opening a pathway toward the convergence of ionic transport and optical control - an emerging paradigm we term iono-photonics.

The isolation of atomically thin two-dimensional (2D) materials has unveiled novel regimes of light-matter interaction, creating unprecedented opportunities for nanoscale photonic control[1–3]. This paradigm has been largely propelled by van der Waals (vdW) materials, where weak interlayer forces enable facile exfoliation and the assembly of designer heterostructures[4,5]. However, the pursuit of environmental resilience and the integration of diverse functionalities - such as ionic transport or ferroelectricity - are increasingly shifting the research frontier toward non-vdW layered materials[6]. These systems, often characterized by stronger ionic or covalent interlayer bonding, offer superior mechanical and thermal robustness, along with the rich physics inherent to complex oxides[7]. The successful exfoliation of non-vdW compounds such as $Fe_2O_3$, $FeS_2$ and $InGaS_3$ demonstrates that cleavability is not exclusive to vdW-bonded structures[8–10].

A particularly intriguing class of non-vdW materials is the honeycomb-layered tellurate oxides ($A_2M_2TeO_6$; A = alkali, M = transition metal)[11]. These compounds are extensively studied as solid electrolytes due to their remarkable ionic conductivity, stemming from inherent structural disorder and high mobility within the alkali metal sublayer[12–14]. While their electrochemical properties are well-characterized, the optical potential of these structurally complex architectures remains largely veiled. Recent observations of strong nonlinear optical responses in related tellurates[15] suggest a promising avenue for optical discovery. $Na_2Zn_2TeO_6$ (NZTO) is an example, known for its high ionic conductivity (up to 9.77 mS cm$^{-1}$)[16]. It features robust $[Zn_2TeO_6]^{2-}$ slabs interleaved with highly mobile, disordered $Na^+$ ions[17]. This structure features a rigid framework combined with weak ionic interlayer bonding. This unique combination suggests that the material has both mechanical resilience and the potential to be exfoliated.

The convergence of high ionic mobility and a rigid crystalline framework within NZTO presents an unexplored frontier for light-matter interaction. In this work, we elucidate the fundamental optical response and lattice dynamics of NZTO and demonstrate its mechanical exfoliation into nanoscale dimensions. Utilizing variable-angle spectroscopic ellipsometry, we determine the anisotropic dielectric tensor, revealing high transparency and significant optical birefringence. Temperature-resolved Raman spectroscopy confirms the stability of the oxide framework under varying thermal conditions. Our results establish the critical link between the domains of superionic conductivity and dielectric photonics, thereby paving the way for emergent iono-photonic functionalities.

$Na_2Zn_2TeO_6$ crystallizes in a P2-type layered structure (Fig. 1a), typically assigned to the hexagonal space group $P6_322$ or related subgroups[18]. The structure consists of $[Zn_2TeO_6]^{2-}$ slabs formed by edge-sharing $ZnO_6$ and $TeO_6$ octahedra arranged in a honeycomb lattice. These slabs are stacked along the *c*-axis and interleaved by disordered $Na^+$ ions, which underpins the material's superionic conductivity[16,17]. The structure's layered morphology and resulting planar anisotropy suggest a viable pathway for mechanical or chemical exfoliation.

The exfoliation of non-vdW solids is governed by specific structural and energetic factors that create preferential cleavage planes. Two primary mechanisms reduce the energy required for cleavage: significant surface energy gain from atomic relaxation on the newly formed 2D surfaces, and the existence of crystallographic planes with a low density of out-of-plane ionic or covalent bonds. The feasibility of exfoliation is quantified by the exfoliation energy ($E_{exf}$), which can be predicted using first-principles calculations such as Density Functional Theory (DFT) to identify suitable candidate materials[19].

To evaluate the prospects for mechanical exfoliation, we compared the calculated $E_{exf}$ of NZTO with established layered materials (Fig. 1b). DFT calculations (see Methods in supplementary material) yield an $E_{exf}$ for NZTO that surpasses that of typical vdW solids, reflecting the stronger ionic character of the interlayer bonding. However, it remains comparable to other exfoliable non-vdW oxides such as $Fe_2O_3$. Crucially, standard DFT calculations assume a static, ordered lattice and thus tend to overestimate the interlayer binding energy in systems with inherent disorder. In NZTO, the $Na^+$ ions are highly mobile and statistically distributed within the interlayer space[16,17]. This inherent frustration of the ionic bonds effectively weakens the electrostatic coupling between the $[Zn_2TeO_6]^{2-}$ slabs, mediating the cleavage process.

Experimentally, we validate the mechanical cleavage of NZTO, as illustrated by the large flake deposited on a Si substrate (Fig. 1c). Moreover, atomic force microscopy (AFM) confirms the isolation of flakes down to thicknesses of approximately 4.3 nm (Fig. 1d), corresponding to a few unit cells. This realization of nanoscale NZTO underscores a viable pathway for integrating complex superionic conductors into quasi-2D architectures, significantly broadening the material space beyond the vdW paradigm. The chemical composition of the bulk NZTO crystal was confirmed using energy-dispersive X-ray spectroscopy and X-ray photoelectron spectroscopy (see Methods and Supplementary note S1 in supplementary material for detailed analysis).

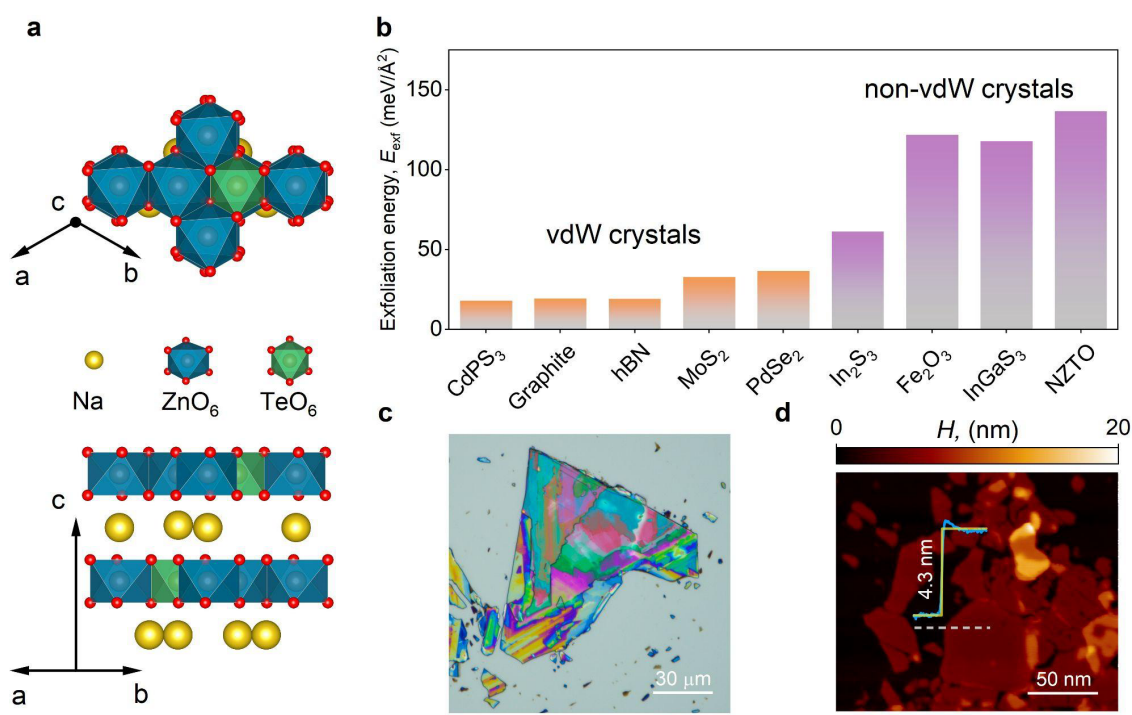

**FIG. 1.** Structure and Exfoliation of $Na_2Zn_2TeO_6$. (a) Crystal structure of P2-type NZTO highlighting the layered stacking of $TeO_6$ and $ZnO_6$ octahedra and the interlayer $Na^+$ ions. (b) Comparison of calculated monolayer exfoliation energies for different layered materials. (c) Optical image of large NZTO flake. (d) AFM image of thin NZTO nanosheets with inset thickness profile showing a 4.3 nm step.

The hexagonal crystal symmetry dictates that NZTO behaves as a uniaxial material, possessing distinct in-plane (ordinary, $n_o$) and out-of-plane (extraordinary, $n_e$) refractive indices. We employed variable-angle spectroscopic ellipsometry (VASE) on a thick exfoliated flake (≈ 1.2 μm) to accurately resolve the anisotropy. The experimental ellipsometric parameters ($\Psi$ and $\Delta$) measured at 40° and 50° angles of incidence are shown in Fig. 2a and 2b. The pronounced Fabry-Perot oscillations indicate high optical quality.

We modeled the data using an anisotropic uniaxial model (see Methods in supplementary material). The analysis confirms that NZTO is essentially lossless (extinction coefficient $k \approx 0$) across the measured spectral range (365–945 nm). This wide transparency window is consistent with a large electronic bandgap ($E_g > 3.6$ eV)[16]. The extracted anisotropic optical constants in comparison with those calculated by DFT Perdew-Burke-Ernzerhof (PBE)[20] are presented in Fig. 2c. (see Methods in supplementary material for detailed description of ab initio computations). The calculated optical constants have a dispersion shape similar to the experimental ones. NZTO exhibits moderate refractive indices, with $n_o$ decreasing from approximately 1.9 at 400 nm to 1.8 at 945 nm.

A salient feature of the optical response is the pronounced optical birefringence ($\Delta n = n_o - n_e \approx 0.25$) sustained across the visible and near-infrared spectrum. This magnitude is remarkable for a wide-bandgap material; it significantly exceeds that of ubiquitous birefringent materials like calcite ($\Delta n \approx 0.17$)[21] and approaches that of rutile (TiO$_2$, $\Delta n \approx 0.29$)[22] in the visible range. The realization of such large optical anisotropy within a readily exfoliable and ionically active oxide architecture positions NZTO as a compelling platform for nanoscale polarization engineering.

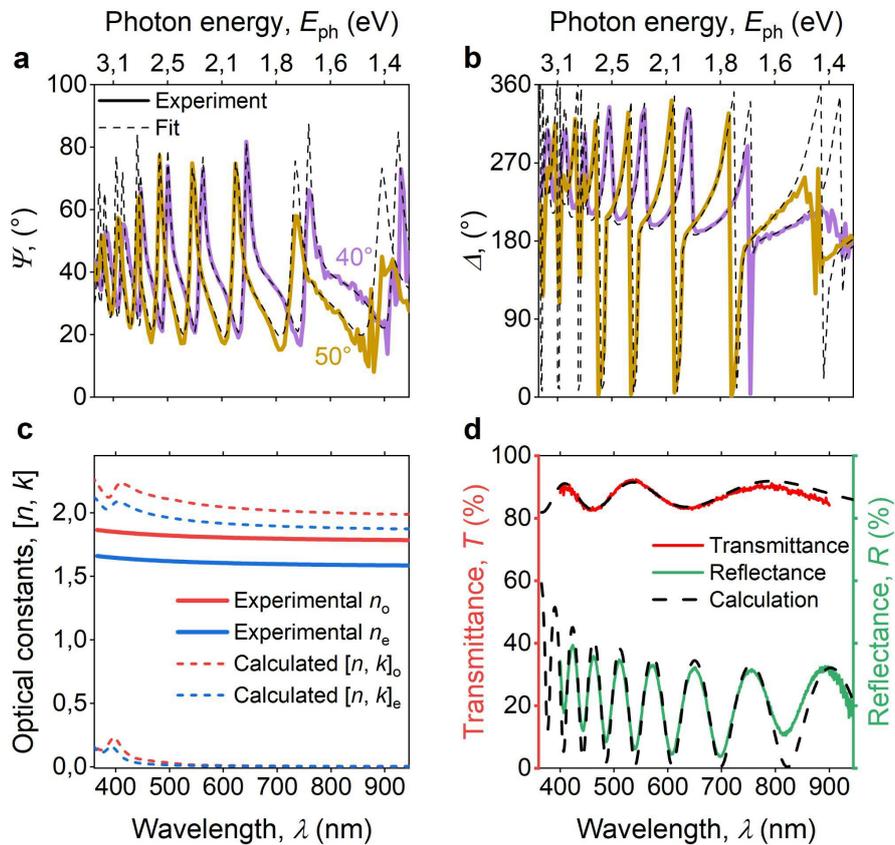

**FIG. 2**. Anisotropic Optical Properties of NZTO. Experimental (solid lines) and fitted (dashed lines) ellipsometric parameters $\Psi$ (a) and $\Delta$ (b) for a 1200 nm thick NZTO flake at incidence angles of 40° (purple) and 50° (gold). (c) Experimentally determined ordinary ($n_o$, red) and extraordinary ($n_e$, blue) refractive indices of NZTO. Dashed lines indicate DFT-PBE calculations. (d) Experimental (solid) and calculated (dashed) transmittance (red/black) and reflectance (green/black) spectra of NZTO flakes.

To validate the ellipsometry-derived constants, we performed normal-incidence micro-reflectance ($R$) and transmittance (*T*) spectroscopy (Fig. 2d). The calculated spectra, generated using the transfer-matrix method with the derived optical constants, show excellent agreement with the experimental data, confirming the accuracy of the determined dielectric tensor.

We utilized Raman spectroscopy to probe the lattice dynamics and structural stability of the framework hosting the mobile ions. The spectrum at 78 K (Fig. 3a) is dominated by a sharp peak at 669 cm$^{-1}$, assigned to the symmetric stretching mode of the TeO$_6$ octahedra[23,24]. Other resolved modes are associated with TeO$_6$ stretching (606 cm$^{-1}$)[18], and lower frequency Na–O or Zn–O dynamics (≈ 260 cm$^{-1}$)[24] (see Tab. S3 for the assignment of all observed Raman modes and Supplementary Note S2 for a detailed analysis of the Raman spectrum in supplementary material).

We investigated the evolution of the vibrational spectra from 78 K to 433 K (Fig. 3b). Upon cooling, all phonon modes exhibit narrow and harden (shift to higher wavenumbers), characteristic of anharmonic phonon-phonon interactions. The behavior of the dominant 669 cm$^{-1}$ mode is analyzed in detail (Fig. 3c, d). As the temperature decreases, the peak position ($x_0$, fitted line center) blue-shifts and the full width at half maximum (FWHM) decreases markedly, reflecting the suppression of phonon damping pathways and reduced thermal lattice expansion.

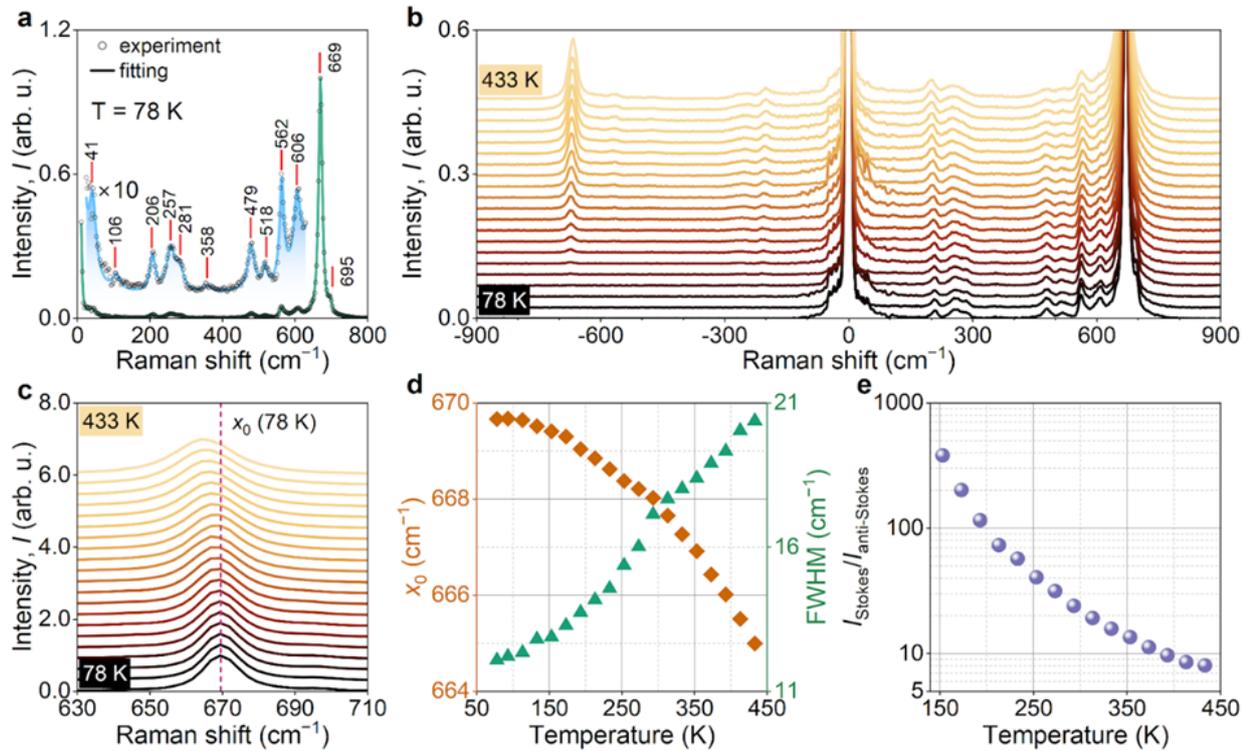

**FIG. 3.** Temperature-dependent Lattice Dynamics of NZTO. (a) Raman spectrum of NZTO at 78 K. Inset shows a magnified low-frequency region. (b) Evolution of Raman spectra (Stokes and anti-Stokes) with temperature (78–433 K). (c) Expanded view of the 669 cm$^{-1}$ peak shift. (d) Temperature dependence of the peak position ($x_0$, orange) and FWHM (green) for the 669 cm$^{-1}$ mode. (e) Temperature dependence of the Stokes-to-anti-Stokes intensity ratio.

The stability of the Raman spectra across this wide temperature range confirms the robustness of the [Zn$_2$TeO$_6$]$^{2-}$ framework. Despite the known high mobility and disorder of the Na$^+$ ions, the vibrational modes governing the material's primary optical response appear largely decoupled from the ionic dynamics in the interlayer space within this frequency range. This robustness is essential for reliable photonic operation. Furthermore, the temperature dependence of the Stokes/anti-Stokes intensity ratio (Fig. 3e) follows the expected Boltzmann distribution, providing a robust method for non-invasive local thermometry.

NZTO's optical properties, including wide transparency, the moderate refractive index, and birefringence, position it as an interesting material for use in integrated polarization optics. These combined characteristics make it highly suitable for manufacturing devices that control light polarization on a chip. We explore this potential by analyzing the interface between NZTO and a high-index P3 polymer[25] (Fig. 4a). Such an interface can function as a polarizing beamsplitter by exploiting the generalized Brewster effect in anisotropic media. A significant advantage of this approach is the ease of producing a real prototype.

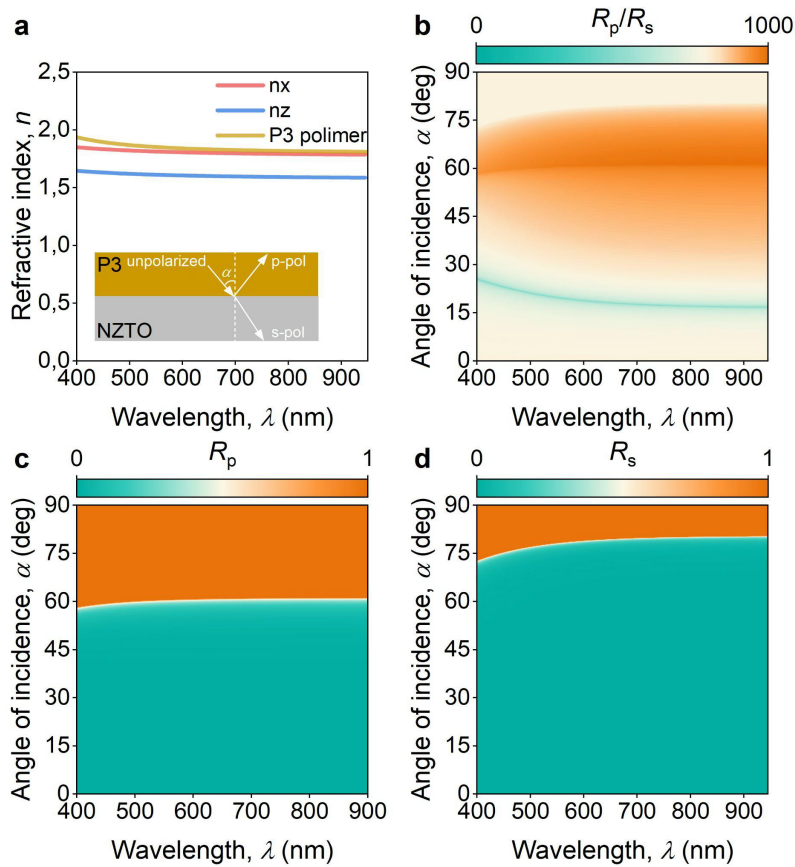

**FIG. 4.** NZTO Interface for Polarization Control. (a) Refractive indices of NZTO ($n_o$, $n_e$) and the P3 polymer. Inset: a schematic illustration of the P3/NZTO interface acting as a polarizing beamsplitter. (b) Calculated extinction ratio ($R_p/R_s$) map. Calculated reflectance maps for (c) p-polarization ($R_p$) and (d) s-polarization ($R_s$).

We calculated the reflectance for p-polarized ($R_p$) and s-polarized ($R_s$) light as a function of wavelength and incidence angle using generalized Fresnel equations (see Supplementary note S3 in supplementary material). The results are shown in Fig. 4c and 4d. The structure exhibits a clear minimum in $R_p$ near the Brewster angle, while $R_s$ remains high across the spectrum. This differential reflectivity results in a strong polarization contrast. The extinction ratio ($R_p/R_s$) map (Fig. 4b) demonstrates that high polarization efficiency can be achieved across a broad spectral range, highlighting the potential for utilizing exfoliated NZTO in robust, integrated polarization control devices. Also, Fig. 4b's map clearly displays a dark orange region where the polarization ratio $R_p/R_s$ reaches 1000. This high ratio is attributed to the close match between the optical constants of the crystal and the polymer, an effect previously shown with a $GeS_2$/$TiO_2$ beamsplitter interface[26]. To maximize performance, one can optimize the polymer mixture to achieve optical constants that are as close as possible to those of NZTO.

In summary, we have unveiled the emergent optical functionality of $Na_2Zn_2TeO_6$, establishing it as a material platform that bridges the realms of superionic conductivity and dielectric photonics. By demonstrating the nanoscale exfoliation of this non-vdW complex oxide - facilitated by its disordered ionic interlayer - we expand the available toolkit of layered materials to include architectures with inherently superior environmental robustness.

Optically, NZTO functions as a wide-bandgap dielectric characterized by low losses and substantial birefringence ($\Delta n \approx 0.25$), providing the foundational parameters for advanced optical design. The most compelling aspect of this platform lies in the coexistence of a rigid optical framework and highly mobile $Na^+$ ions. This synergy defines a new frontier in iono-photonics, offering a distinct pathway for realizing active, reconfigurable optical devices. Future efforts will focus on harnessing electrostatic gating and ionic intercalation to dynamically manipulate the optical response of NZTO, leveraging its unique material properties to control light propagation at the nanoscale.

## Supplementary Material

Supplementary Material contains sections Methods, Supplementary notes S1, S2 and S3.

## Author Contributions

A.N.T., G.A.E., A.V.A., and V.S.V. suggested and directed the project. M.K.T., A.S.S., N.V.P., M.A.E.-S., A.A.M. and D.I.Y. performed the measurements and analyzed the data. I.A.K and K.V.K. performed ab initio calculations. D.V.G. and A.A.V. provided theoretical support. N.V.P. wrote the original manuscript. All authors reviewed and edited the paper. All authors contributed to the discussions and commented on the paper.

## Competing Interests

The authors declare no competing interests.

## Acknowledgments

N.V.P., M.K.T., K.V.K. and A.V.A. acknowledge the financial support from the RSF (Grant No. 25-19-00326).

## Data Availability

The datasets generated during and/or analyzed during the current study are available from the corresponding author upon reasonable request.

# References

[1] A.K. Geim, and K.S. Novoselov, "The rise of graphene," Nat. Mater. 6(3), 183–191 (2007).

[2] Wallace, P. R. "The Band Theory of Graphite," Physical Review B 71(9), 622–634 (1947).

[3] K.F. Mak, and J. Shan, "Photonics and optoelectronics of 2D semiconductor transition metal dichalcogenides," Nat. Photonics 10(4), 216–226 (2016).

[4] A.K. Geim, and I.V. Grigorieva, "Van der Waals heterostructures," Nature 499, 419–425 (2013).

[5] H. Xu, Y. Xue, Z. Liu, Q. Tang, T. Wang, X. Gao, Y. Qi, Y.P. Chen, C. Ma, and Y. Jiang, "Van der Waals heterostructures for photoelectric, memory, and neural network applications," Small Sci. 4(4), 2300213 (2024).

[6] I. Tantis, S. Talande, V. Tzitzios, G. Basina, V. Shrivastav, A. Bakandritsos, and R. Zboril, "Non-van der Waals 2D Materials for Electrochemical Energy Storage," Adv. Funct. Mater. 33(19), (2023).

[7] R. Friedrich, M. Ghorbani-Asl, S. Curtarolo, and A.V. Krasheninnikov, "Data-Driven Quest for Two-Dimensional Non-van der Waals Materials," Nano Lett. 22(3), 989–997 (2022).

[8] A.N. Toksumakov, G.A. Ermolaev, A.S. Slavich, N.V. Doroshina, E.V. Sukhanova, D.I. Yakubovsky, A.V. Syuy, S.M. Novikov, R.I. Romanov, A.M. Markeev, A.S. Oreshonkov, D.M. Tsymbarenko, Z.I. Popov, D.G. Kvashnin, A.A. Vyshnevyy, A.V. Arsenin, D.A. Ghazaryan, and V.S. Volkov, "High-refractive index and mechanically cleavable non-van der Waals $InGaS_3$," Npj 2D Mater. Appl. 6(1), (2022).

[9] J. Mohapatra, A. Ramos, J. Elkins, J. Beatty, M. Xing, D. Singh, E.C. La Plante, and J. Ping Liu, "Ferromagnetism in 2D α-$Fe_2O_3$ nanosheets," Appl. Phys. Lett. 118(18), (2021).

[10] A.B. Puthirath, A.P. Balan, E.F. Oliveira, V. Sreepal, F.C. Robles Hernandez, G. Gao, N. Chakingal, L.M. Sassi, P. Thibeorchews, G. Costin, R. Vajtai, D.S. Galvao, R.R. Nair, and P.M. Ajayan, "Apparent ferromagnetism in exfoliated ultrathin pyrite sheets," J. Phys. Chem. C Nanomater. Interfaces 125(34), 18927–18935 (2021).

[11] M.A. Evstigneeva, V.B. Nalbandyan, A.A. Petrenko, B.S. Medvedev, and A.A. Kataev, "A new family of fast sodium ion conductors: $Na_2M_2TeO_6$ (M = Ni, Co, Zn, Mg)," Chem. Mater. 23(5), 1174–1181 (2011).

[12] J.-F. Wu, Q. Wang, and X. Guo, "Sodium-ion conduction in $Na_2Zn_2TeO_6$ solid electrolytes," J. Power Sources 402, 513–518 (2018).

[13] F. Bianchini, H. Fjellvåg, and P. Vajeeston, "Nonhexagonal Na sublattice reconstruction in the super-ionic conductor $Na_2Zn_2TeO_6$: Insights from ab initio molecular dynamics," J. Phys. Chem. C Nanomater. Interfaces 123(8), 4654–4663 (2019).

[14] Y. Li, Z. Deng, J. Peng, E. Chen, Y. Yu, X. Li, J. Luo, Y. Huang, J. Zhu, C. Fang, Q. Li, J. Han, and Y. Huang, "A P2-type layered superionic conductor Ga-doped $Na_2Zn_2TeO_6$ for all-solid-state sodium-ion batteries," Chemistry 24(5), 1057–1061 (2018).

[15] X. Du, X. Guo, Z. Gao, F. Liu, F. Guo, S. Wang, H. Wang, Y. Sun, and X. Tao, "$Li_2MTeO_6$ (M=Ti, Sn): Mid-infrared nonlinear optical crystal with strong second harmonic generation response and wide transparency range," Angew. Chem. Int. Ed Engl. 60(43), 23320–23326 (2021).

[16] H. Huang, Y. Yang, C. Chi, H.-H. Wu, and B. Huang, "Phase stability and fast ion transport in P2-type layered $Na_2X_2TeO_6$(X = Mg, Zn) solid electrolytes for sodium batteries," J. Mater. Chem. A Mater. Energy Sustain. 8(43), 22816–22827 (2020).

[17] F.S. Hempel, W.A. Sławiński, B. Arstad, and H. Fjellvåg, "Superstructure of locally disordered $Na_2Zn_2TeO_6$," Chem. Mater. 36(22), 11084–11094 (2024).

[18] V. Kumar, A. Gupta, and S. Uma, "Formation of honeycomb ordered monoclinic $Li_2M_2TeO_6$ (M = Cu, Ni) and disordered orthorhombic $Li_2Ni_2TeO_6$ oxides," Dalton Trans. 42(42), 14992–14998 (2013).

[19] G. Kresse, and J. Hafner, "Ab initio molecular dynamics for liquid metals," Phys. Rev. B Condens. Matter 47(1), 558–561 (1993).

[20] G. Kresse, and J. Furthmüller, "Efficient iterative schemes for ab initio total-energy calculations using a plane-wave basis set," Phys. Rev. B Condens. Matter 54(16), 11169–11186 (1996).

[21] G. Ghosh, "Dispersion-equation coefficients for the refractive index and birefringence of calcite and quartz crystals," Opt. Commun. 163(1-3), 95–102 (1999).

[22] J. R. DeVore, "Refractive indices of rutile and sphalerite," Journal of the Optical Society of America, *41*(6), 416–419 (1951).

[23] R.L. Frost, "Tlapallite $H_6(Ca,Pb)_2(Cu,Zn)_3SO_4(TeO_3)_4TeO_6$, a multi-anion mineral: A Raman spectroscopic study," Spectrochim. Acta A Mol. Biomol. Spectrosc. 72(4), 903–906 (2009).
[24] C. Julien, "Raman spectra of birnessite manganese dioxides," Solid State Ion. 159(3-4), 345–356 (2003).
[25] J. Zhang, T. Bai, W. Liu, M. Li, Q. Zang, C. Ye, J.Z. Sun, Y. Shi, J. Ling, A. Qin, and B.Z. Tang, "All-organic polymeric materials with high refractive index and excellent transparency," Nat. Commun. 14(1), 3524 (2023).
[26] A.S. Slavich, G.A. Ermolaev et al., "Germanium disulfide as an alternative high refractive index and transparent material for UV-visible nanophotonics," Light Sci. Appl. 14(1), 213 (2025).